\documentclass[usegraphicx,useAMS,usenatbib]{mn2e}
\usepackage{times}
\usepackage{amssymb}
\usepackage{amsmath}
\usepackage{graphicx}
\usepackage{euscript}
\usepackage{rotating}
\usepackage{epsfig}
\usepackage{url}

\def\apjl{ApJ}                   

\def\mhz{{\rm\thinspace MHz}}
\def\ghz{{\rm\thinspace GHz}}

\def\kev{{\rm\thinspace keV}}

\newcommand{\eg}{e.g.\thinspace}

\begin{document}

\title{AGN-Induced Cavities in NGC\,1399 and NGC\,4649}
\author[K. Shurkin et al.]
{\parbox[]{6.in} {K. Shurkin$^1$, R.J.H. Dunn$^2$, G. Gentile$^1$, G.B. Taylor$^1$, and S.W. Allen$^3$ \\
    \footnotesize
    $^1$University of New Mexico, Department of Physics and Astronomy, Albuquerque, NM 87131, USA. \\
    $^2$School of Physics and Astronomy, University of Southampton, Southampton, SO17 1BJ, UK. \\
    $^3$Kavli Institute for Particle Astrophysics and Cosmology, Stanford University, 382 Via Pueblo 
        Mall, Stanford, CA 94305-4060, USA. \\
  }}

\maketitle

\begin{abstract}
We present an analysis of archival \textit{Chandra} and VLA observations of the E0 galaxy NGC\,1399 
and the E2 galaxy NGC\,4649 in which we investigate cavities in the surrounding X-ray emitting medium 
caused by the central AGN. We calculate the 
jet power required for the AGN to evacuate these cavities and find values of 
$\sim 8\times 10^{41}$ erg s$^{-1}$ and $\sim 14\times 10^{41}$ erg s$^{-1}$ for the lobes of 
NGC\,1399 and 
$\sim 7\times 10^{41}$ erg s$^{-1}$ and $\sim 6\times 10^{41}$ erg s$^{-1}$ for those of NGC\,4649. 
We also calculate the $k/f$ values for each cavity,
where $k$ is the ratio of the total particle energy to that of electrons radiating in the range of 10 
MHz to 10 GHz, and $f$ is the volume filling factor of the plasma in the cavity. We find that the
values of $k/f$ for the lobes of NGC\,1399 are $\sim93$ and $\sim190$, and those of the lobes of
NGC\,4649 are $\sim15000$ and $\sim12000$. We conclude that the assumed spectrum describes the
electron distribution in the lobes of NGC\,1399 reasonably well, and that there are few entrained
particles. For NGC\,4649, either there are many entrained particles or the model spectrum does not 
accurately describe the population of electrons.  
\end{abstract}


\section{Introduction}\label{sec:intro}
Within nearby galaxies, groups, and clusters, embedded active galactic nuclei (AGN) have been found to
interact with the surrounding hot, X-ray emitting thermal gas, causing such disturbances as shocks,
ripples, and cavities (\eg \citealp{Fabian00, Fabian03b, Fabian06, Forman05}). 
 The cavities, which 
appear as X-ray surface brightness depressions, consist of low density relativistic plasma and are a 
consequence of the thermal gas being displaced by the jets of the AGN. They have been found to be 
associated with the central AGN of many clusters, such as Perseus (\citealp{Bohringer93,Fabian00, 
Fabian03b, Fabian06}), Hydra A (\citealp{McNamara00}), M\,87 (\citealp{Churazov01, Forman05}), and
Centaurus (\citealp{Taylor02,Taylor06a}). AGN-induced disturbances are also present within the hot 
interstellar medium (ISM) in the haloes of normal elliptical galaxies \citep{Diehl06a}. Many early-type 
galaxies, such as M\,84 (\citealp{Finoguenov01}), have been found to harbor cavities in their X-ray 
haloes. Our interest in these AGN-induced cavities stems from the fact that they can be used as 
calorimeters. The kinetic energy of the jets emanating from the central black hole can be estimated by 
calculating the amount of energy required to inflate the observed cavities (\citealp{Birzan04, 
Rafferty06, Dunn06f}). Using this estimate, the jet power can be calculated once an approximate age of 
the cavity has been deduced.

For nearby large, elliptical galaxies, accurate measurements of the density and temperature profiles 
of the thermal gas can be obtained from high resolution \textit{Chandra} observations. For those systems 
that are sufficiently close, such that radii within an order of magnitude of the accretion radius are 
resolved, these measurements can be used to calculate the Bondi accretion rate, given estimates of the 
black hole mass. \citet{Allen06}, in a study of 9 X-ray luminous sources that harbor cavities, found 
that a tight physical correlation exists between the power available from Bondi accretion of the hot 
gas and the observed jet power. These results are significant, especially for studies of accretion and 
jet formation, as well as the formation and evolution of galaxies. In particular, building upon earlier 
studies by \citet{DiMatteo03} and \citet{Taylor06a}, they confirmed that the bulk of the energy produced 
by the central AGN in these systems, which are all \citet{Fanaroff74} type-I sources, is released in the 
relativistic jets. The energy injected back into the environment is in principle sufficient to balance 
radiative cooling of the hot X-ray gas. This possibility has also been investigated by \citet{Best05, 
Best06b}, who showed that the cooling of hot gas from the atmosphere of the host galaxy can feasibly 
provide the fuel for low power, radio-loud AGN, and that the heating by AGN feedback can balance this 
cooling, except perhaps in the most massive clusters. However, the manner in which the energy provided 
by the AGN is coupled to the thermal energy of the hot gas in order to offset cooling, as well as
the efficiency of this process, are as yet unknown.

Because of the significance of AGN activity, it is desirable to investigate further the population of 
galaxies which exhibit interactions between the central AGN and the surrounding medium. In particular, we 
are interested in those sources which harbor cavities, as they offer insight into the physical properties 
of the AGN itself. 

We analyzed \textit{Chandra} and VLA\footnote{The VLA (Very Large Array) is operated by the National Radio 
Astronomy Observatory (NRAO).} data for NGC\,1399 and NGC\,4649, both of which are large, elliptical 
galaxies that harbor cavities and are located in nearby clusters. Because of their proximity, we are able
to determine the density of the X-ray emitting gas down to radii within a factor of 10 of the accretion 
radius, making these two galaxies ideal for future analysis concerning Bondi accretion. At present, we use 
the cavities in these systems to estimate the kinetic energy of the jets and jet power. We also investigate 
the particle content of the cavities by determining $k/f$, where $k$ is the ratio of the total particle 
energy contained in the cavity to the energy accounted for by electrons emitting synchrotron radiation in 
the range of 10 MHz to 10 GHz, and $f$ is the volume filling factor of the relativistic plasma in the cavity.
The cavities are filled with relativistic charged particles and
magnetic fields, which together determine the synchrotron emission from
the lobes. The relative energy densities of these two components, and
hence $k$ and $f$, cannot be disentangled from the emission alone. We must
therefore consider the ratio $k/f$, which, under the assumption that the
gas pressure is equal to the total pressure from the radio plasma, can
be related to the magnetic field strength.  The maximum
field strength is obtained for the minimum value of $k/f=1$, and would
indicate a pure electron-positron plasma uniformly filling the cavity. 

Our discussion will proceed as follows. We begin in \S \ref{sec:radio} by describing the radio data selection 
and processing. In \S \ref{sec:xray}, we describe the X-ray data and preparation, as well as the 
identification of cavities. In \S \ref{sec:interaction} we present \textit{Chandra}/VLA comparisons for
NGC\,1399 and NGC\,4649 and determine the power of the jets in these systems. We present the $k/f$ ratios in 
\S \ref{sec:kf}, followed by conclusions in \S \ref{sec:concl}. 

Throughout this discussion, we assume a flat $\Lambda$CDM cosmology with $\Omega_M=0.3$ and 
$H_0=70$ km s$^{-1}$ Mpc$^{-1}$.

\section{Radio Data: Reduction and Imaging}\label{sec:radio}

The VLA radio data were obtained from the NRAO archive. In selecting the data to be analyzed, preference was given 
to observations with time on source $\geq 5$ minutes. Because steep spectrum emission is brighter at lower 
frequencies, preference was also given to observations performed at 1.4~GHz. In addition, A-configuration or 
B-configuration observations at 1.4~GHz were desirable in order to provide arcsecond resolution for 
comparison with X-ray images from \textit{Chandra}. Parameters for the VLA observations of NGC\,1399 and
NGC\,4649 are tabulated in Table \ref{tab:obs}.

The radio data were reduced in the standard manner using {\scshape aips}. After an initial editing of the 
data, absolute amplitude and phase calibration were performed on each dataset using the scripts {\scshape 
vlaprocs} and {\scshape vlarun}. As the flux calibrator for each dataset was resolved, a model was 
used for the calibration. 
If bad data were still present after the initial calibration, those 
data were flagged and the calibration was repeated.
 
Each source was imaged using the task {\scshape imagr}. Sidelobes from outlying sources were removed by using 
multiple facets while imaging. Proper placement of the facets was determined using the task {\scshape setfc}, 
which was set up to search a 0.5 degree radius for sources in the NVSS\footnote{The NVSS (NRAO VLA Sky Survey) 
is a 1.4 GHz continuum survey covering the entire sky north of -40$^o$ declination \citep{Condon98}.} 
catalogue with flux $\geq 10$ mJy. Typically, sources outside of that range are either not bright enough or 
too far from the pointing center to have an appreciable effect on the quality of the image.
As both sources had 
sufficient signal-to-noise ratios, imaging and phase-only self-calibration were then performed iteratively, 
until the theoretical noise was reached or until the quality of the map ceased to benefit from the iterations. 
Radio flux density measurements are shown in Table \ref{tab:vla}.

\begin{table*}
\caption{Positions, redshifts, and radio and X-ray observation details for NGC\,1399 and NGC\,4649. The 
frequency (in GHz), configuration, date, and duration (in s) of the VLA observations are shown in columns 5
through 8. The ObsID and final exposure time (in ks) of the \textit{Chandra} observations
are shown in columns 9 and 10.\label{tab:obs}}
\begin{tabular}{lccccccccccc}
\hline
\hline
\ & \ & \ & \ & \ & \multicolumn{4}{c}{\textsc{VLA}} & \ & \multicolumn{2}{c}{\textsc{Chandra}}\\
\hline
Source & RA$^1$ & Dec$^1$ & Redshift$^1$ & \ & Frequency & Config & Date & Duration & \ &  ObsID & Exposure$^2$\\
\ & (J2000) & (J2000) & \ & \ & (\ghz) & \ & \ & (sec) & \ & \ & (ks)\\
\hline

NGC\,1399 & 03h\,38m\,29.08s & $-$35d\,27m\,02.7s & 0.004753 & \ &  1.465 & H$^3$ & 1983 Dec 16 & 27910 & \ & 319 & 55.9\\
NGC\,4649 & 12h\,43m\,39.66s & +11d\,33m\,09.4s & 0.003726 & \ & 1.465 & B & 1984 Jan 24 & 4410 & \ & 785 & 36.9\\

\hline
\end{tabular}
\begin{quote}
\

$^1$Positions and redshifts obtained from the NASA/IPAC Extragalactic Database (NED).

$^2$The exposure after reprocessing.

$^3$Hybrid configuration (A and B).
\end{quote}
\end{table*}

\begin{table*}
\caption{Radio flux density measurements\label{tab:vla}}
\begin{tabular}{lccccc}
\hline
\hline
Source & Peak Flux & Total Flux Density & RMS & Beam & PA \\
\ & (Jy beam$^{-1}$) & (Jy) & ($\mu$Jy beam$^{-1}$) & (arcsec$^2$) & (deg) \\
\hline
NGC\,1399 & 1.89$\times10^{-2}$ & 4.63$\times10^{-1}$ & 72.1 & 4.15$\times$2.80 & 38.5 \\
NGC\,4649 & 1.72$\times10^{-2}$ & 2.82$\times10^{-2}$ & 29.7 & 4.51$\times$3.64 & 44.8 \\
\hline
\end{tabular}
\end{table*}

\section{X-ray Data Preparation}\label{sec:xray}

The {\it Chandra} X-ray data were used to determine the deprojected temperature and density profiles of the
X-ray emitting gas in the galaxies. These profiles, in 
combination with the measured sizes of the cavities identified from the X-ray and radio emission, allow the 
$P\,\textrm{d}V$ work done by the cavity on the surrounding gas halo to be calculated. The details of the 
X-ray observations are summarized in Table \ref{tab:obs}.

Before analysis, the X-ray data were reprocessed and cleaned using the {\scshape ciao} software and calibration files
({\scshape ciao} v3.3, {\scshape caldb} v3.2). We began the reprocessing by removing the afterglow detection 
and re-identifying the hot pixels and cosmic ray afterglows, followed by the tool 
{\scshape acis\_process\_events} to remove the pixel randomization and to flag potential background events 
for data observed in Very Faint (VF) mode.  The Charge-Transfer Inefficiency (CTI) was corrected for, followed by
standard grade selection.  Point sources were identified using the {\scshape wavdetect} wavelet-transform 
procedure. As these two galaxies were observed with the ACIS-S3 chip, background lightcurves to check for 
flaring were taken from the ACIS-S1 chip. For the spectral analysis, backgrounds were taken from the CALDB
blank-field datasets.  They had the same reprocessing applied, and were reprojected to the correct 
orientation.  

Spectra were extracted in annular regions centered on the peaks of the X-ray emission from the galaxies. 
Annular regions were automatically assigned with constant signal-to-noise, stopping where the background 
subtracted surface brightness became consistent with zero. The initial signal-to-noise was 100, and this was
increased or decreased by successive factors of $\sqrt{2}$ to obtain a number of regions between four and ten.  
The minimum signal-to-noise allowed was 10. 

The $0.5-7\kev$ spectra were extracted, binned with a minimum of 20 counts per bin, and, using {\scshape xspec} 
(v12.3.0) (e.g. \citealt{Arnaud96}), a {\scshape projct} single temperature {\scshape mekal} 
(e.g. \citealt{Mewe95}) model with a {\scshape phabs} absorption was used to deproject the cluster. The 
deprojected cluster temperature, abundance and normalization profiles allowed the calculation of density and 
pressure profiles.  Chi-square was used as the estimator in the fits.

These profiles give azimuthally averaged values for the cluster properties and have been used in the subsequent 
calculations. 
The substructure present in these galaxies is not expected to bias the results on the gas properties 
significantly \citep{Donahue06}. As such the use of these azimuthally averaged values is not likely to 
introduce large biases into the subsequent calculations.

\section{Interaction Between AGN Jets and Thermal Gas}\label{sec:interaction}

\begin{figure*}
\centering
$\begin{array}{l@{\hspace{0.05in}}l}
\epsfxsize=3.40in
\epsffile{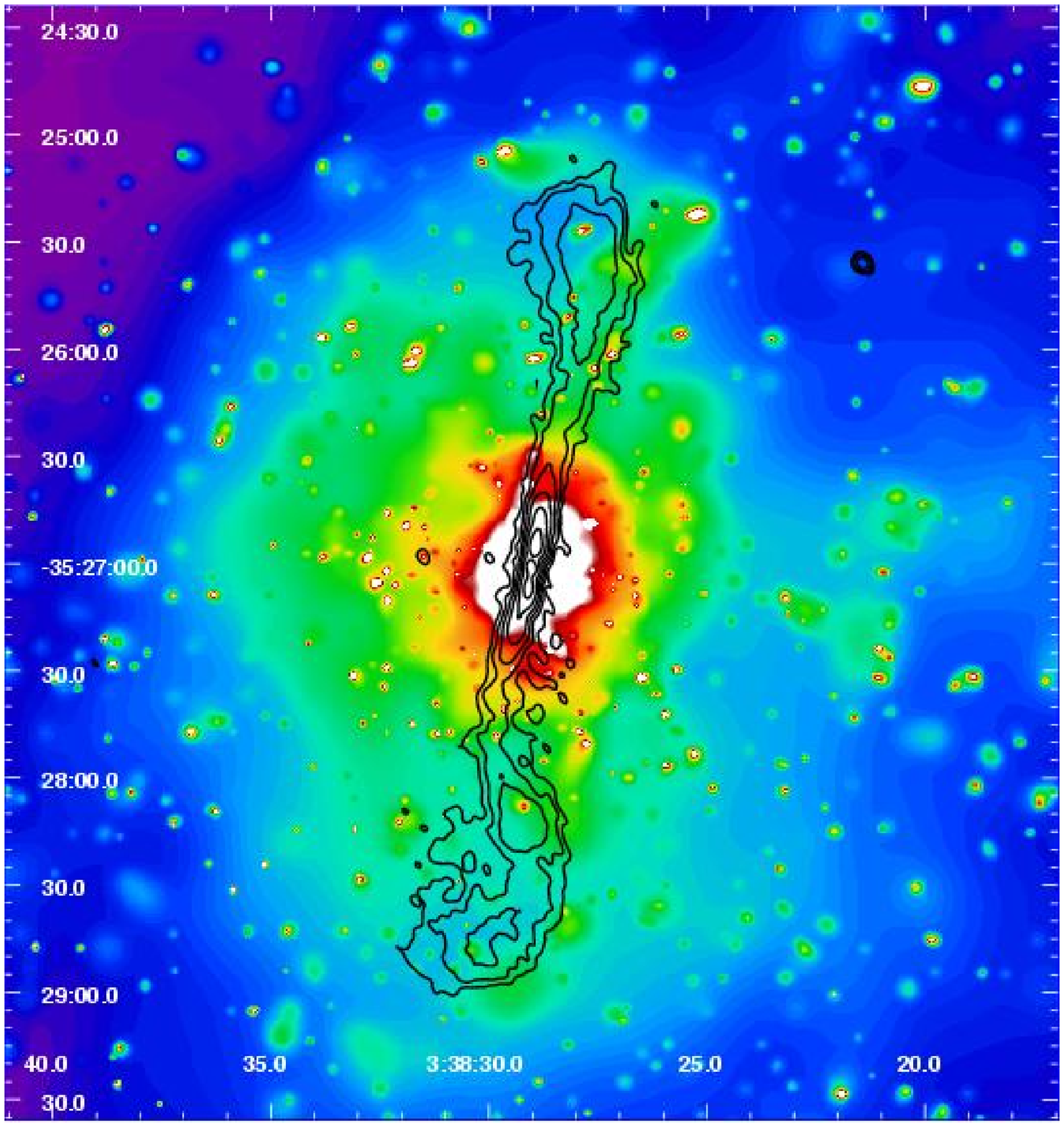} &
          \epsfxsize=3.40in
          \epsffile{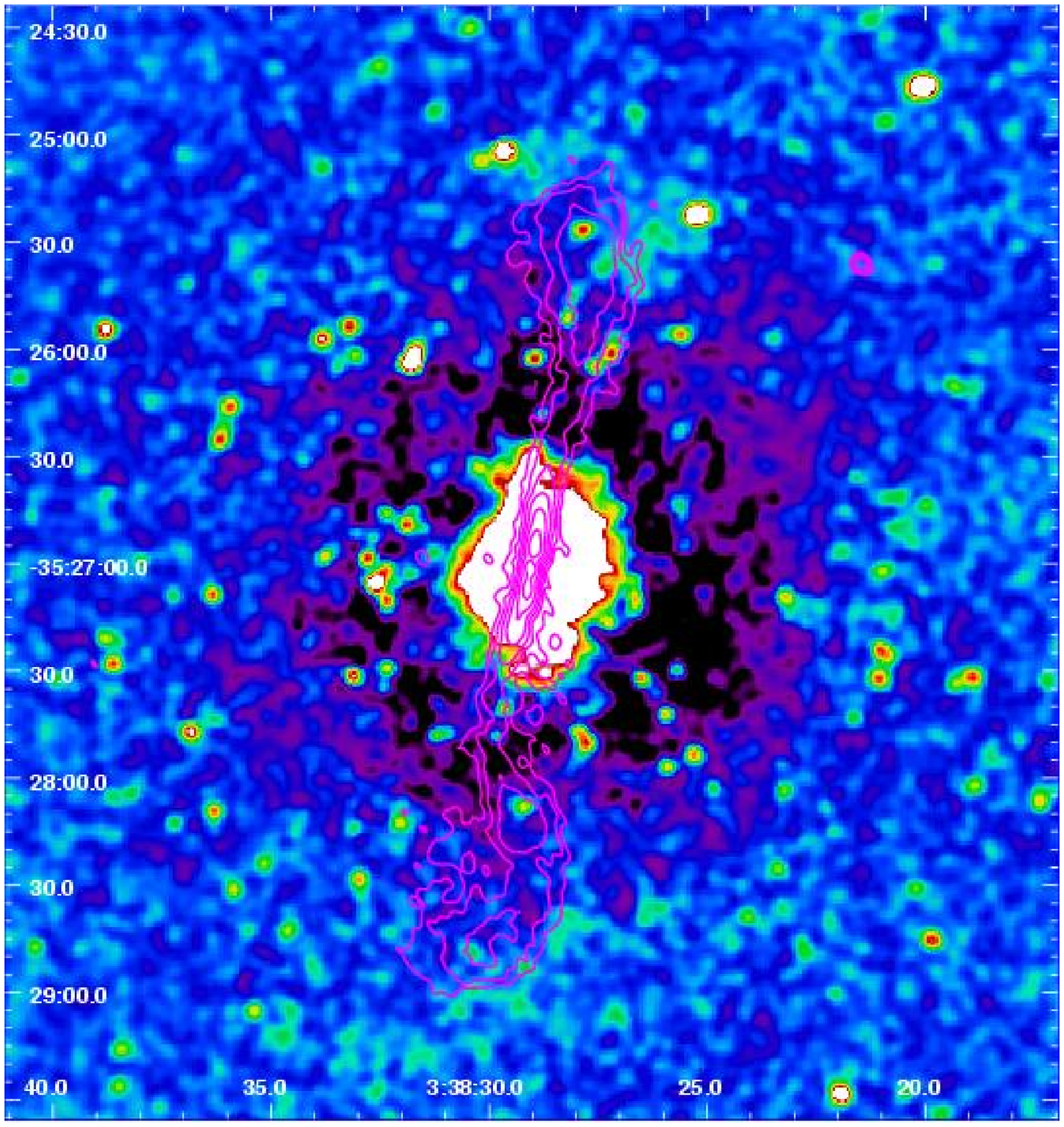} \\
\multicolumn{1}{c}{\mbox{\bf (a)}} & \multicolumn{1}{c}{\mbox{\bf (b)}} 
\end{array}$
\caption{X-ray images of NGC\,1399 with 1.4 GHz radio contours. The left panel 
  shows an adaptively smoothed X-ray image with a logarithmic color scale, and the right panel shows an 
  unsharp-masked X-ray image with a linear color scale. 
  The unsharp-masking was done by subtracting a raw image smoothed using an 80 arcsec Gaussian kernel from one 
  smoothed using a 4 arcsec Gaussian kernel. The dark ring in the unsharp-masked image is
  due to the central source being much brighter than the rest of the image.}
\label{fig:1399}
\end{figure*}

\begin{figure*}
\centering
$\begin{array}{l@{\hspace{0.05in}}l}
\epsfxsize=3.40in 
\epsffile{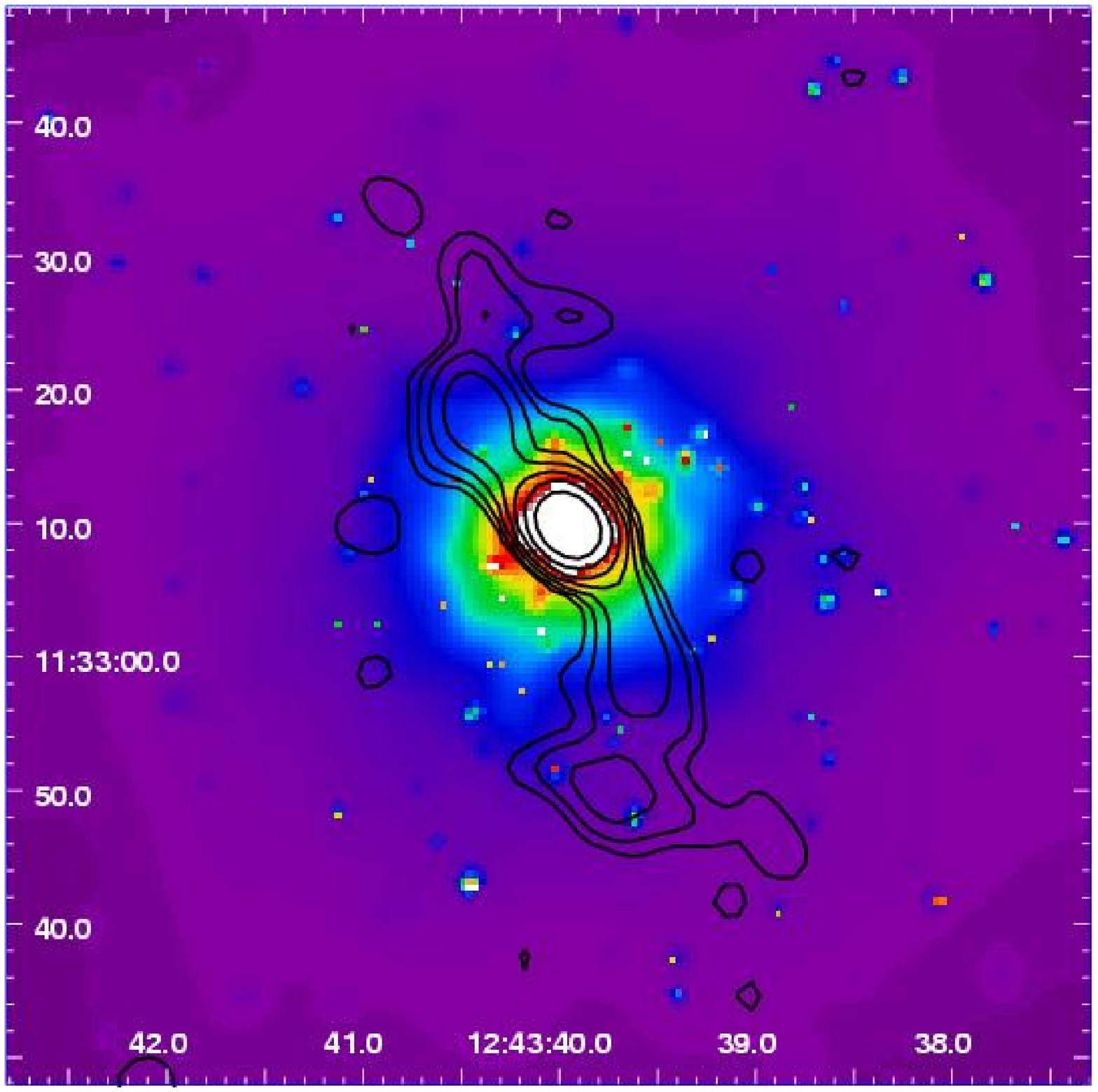} &
          \epsfxsize=3.40in
          \epsffile{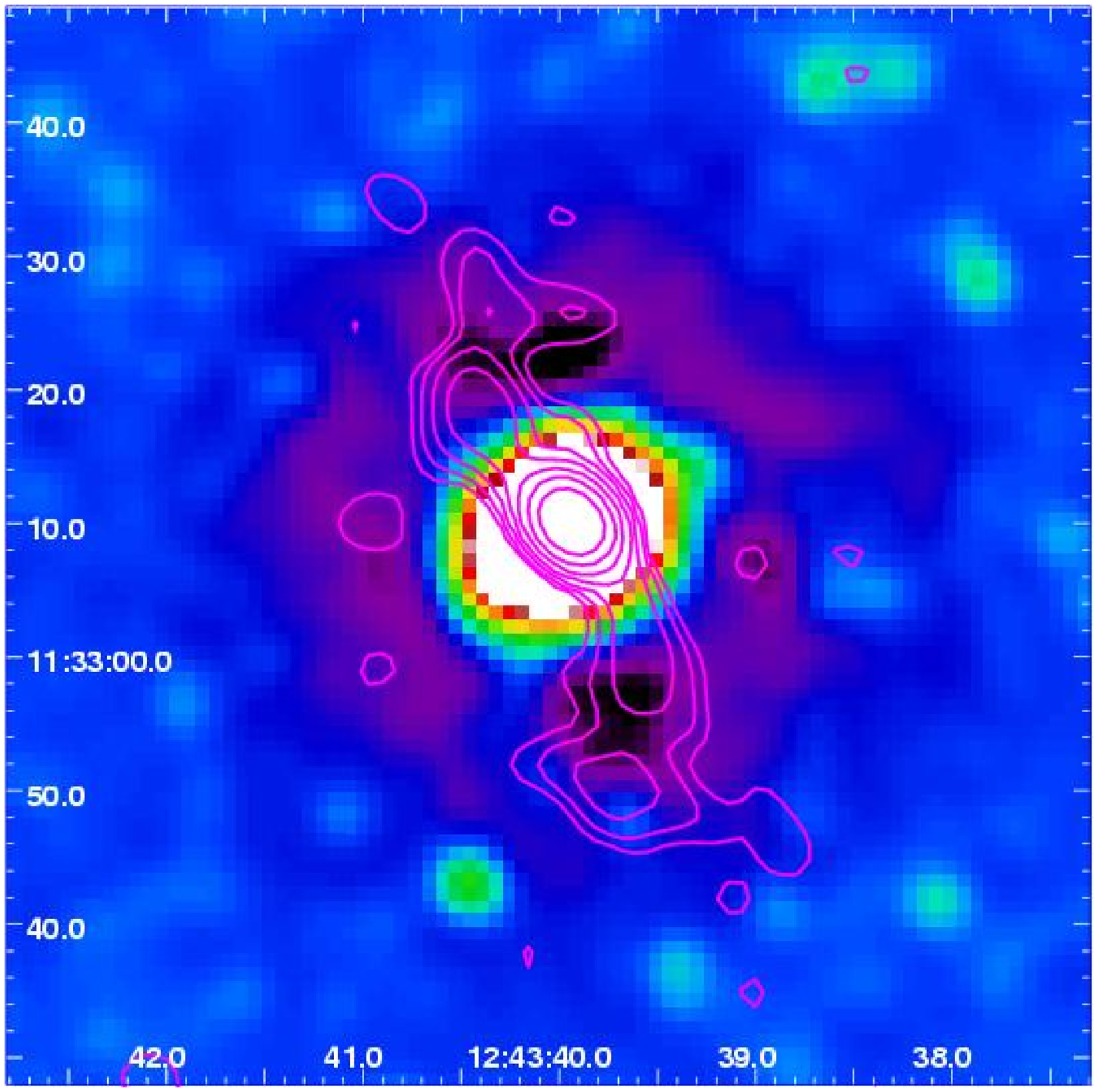} \\
\multicolumn{1}{c}{\mbox{\bf (a)}} & \multicolumn{1}{c}{\mbox{\bf (b)}}
\end{array}$
\caption{X-ray images of NGC\,4649 with 1.4 GHz radio contours. The left panel 
  shows an adaptively smoothed X-ray image with a logarithmic color scale, and the right panel shows 
  an unsharp-masked X-ray image with a linear color scale.
  The unsharp-masking was done by subtracting a raw image smoothed using a 20 arcsec Gaussian kernel from one 
  smoothed using a 4 arcsec Gaussian kernel.}
\label{fig:4649}
\end{figure*}

In order to investigate any possible interactions between the AGN and the surrounding thermal gas, we 
superimposed the VLA radio contours onto the \textit{Chandra} X-ray images. These X-ray/radio overlays are 
shown in Figures \ref{fig:1399} and \ref{fig:4649}. Panels (a) and (b) of each figure show an adaptively
smoothed X-ray 
image and an unsharp-masked X-ray image, respectively, with color scales chosen to highlight faint features.
The adaptive smoothing was done with a 2 arcsec kernel
up to a maximum of 20 using {\scshape asmooth} \citep{Ebeling06}. The unsharp-masked images are difference 
images made by first smoothing the raw image using a large Gaussian kernel (80 arcsec for NGC\,1399 and 20 
arcsec for NGC\,4649), and then subtracting that image from one which has been smoothed using a 4 arcsec
Gaussian kernel. The radio contours aid in the 
determination of the approximate size and location of the cavities. We use these estimates, in conjunction with 
density and temperature profiles derived from the X-ray data, to estimate the jet power required to inflate the 
observed cavities. These calculations are summarized in \S \ref{sub:pjet}, following individual discussions of  
NGC\,1399 and NGC\,4649 in \S \ref{sub:1399} and \S \ref{sub:4649}, respectively. Some additional derived 
properties of the galaxies are presented in Table \ref{tab:properties}. 

\begin{table}
\caption{Properties\label{tab:properties}}
\begin{tabular}{lll}
\hline
\hline
\ & NGC\,1399 & NGC\,4649 \\
\hline
Distance (Mpc) & 19.4 & 17.1 \\
$L_{\rm x}$ (erg s$^{-1}$)$^{\star}$ & $4.27\times10^{41}$ & $1.91\times10^{41}$ \\
$P_{1.4}$ (erg s$^{-1}$ Hz$^{-1}$)$^{\dagger}$ & $2.09\times10^{29}$ & $9.87\times10^{27}$ \\
Scale (kpc arcmin$^{-1}$) & 5.64 & 4.96 \\
Largest angular size (arcsec)$^{\ddagger}$ & 233 & 46.5 \\
Largest physical size (kpc)$^{\ddagger}$ & 21.9 & 3.85 \\
\hline
\end{tabular}
\begin{quote}
\

$^{\star}$ X-ray luminosity, from \citet{OSullivan01}.

$^{\dagger}$Total isotropic radio power at 1.4~GHz: $P_{1.4}=4\pi D^2S_{1.4}$.

$^{\ddagger}$Radio source size.
\end{quote}
\end{table}

\subsection{NGC\,1399}\label{sub:1399}

NGC\,1399 is the central E0 galaxy of the Fornax cluster. It is host to a low power radio source, whose lobes are 
confined within the optical galaxy. \citet{Paolillo02} present a ROSAT/VLA comparison which shows interactions 
between the central AGN and the surrounding hot gas. Figure \ref{fig:1399} shows our \textit{Chandra}/VLA 
comparison for NGC\,1399. These comparisons confirm the structures found in \citet{Paolillo02}, as well as the 
presence of cavities associated with the radio lobes. The lobes extend north and south along an axis oriented 
$\sim$10 degrees west of north. They are approximately symmetric, each having a length of roughly 11 kpc. The 
central X-ray source is elongated in the direction of the lobes, with bulges along the east and west edge of 
the northern and southern lobes, respectively (see Figure \ref{fig:1399}b). The southern lobe bends sharply 
west before gradually bending eastward. As can be seen in Figure \ref{fig:1399}(a), a bright rim of X-ray 
emission follows the western edge of the southern lobe, marking the edge of the cavity. The northern lobe also 
bends slightly eastward at the end, and a bright rim of emission can be seen in Figure \ref{fig:1399}(a) to 
follow the western edge of this lobe as well. Both lobes end in a bright clump of X-ray emission.

\subsection{NGC\,4649}\label{sub:4649}

NGC\,4649, also known as M\,60, is an E2 galaxy located in a group at the eastern edge of the Virgo cluster. 
Figure \ref{fig:4649} shows our \textit{Chandra}/VLA comparison for NGC\,4649. Figure \ref{fig:4649}(b) shows 
cavities in the X-ray emission that are coincident with the lobes of the weak radio source. The lobes extend 
northeast and southwest, along an axis approximately 31 degrees east of north. The northern lobe bends westward, 
while the southern lobe bends eastward, giving the radio source an overall S shape. The central X-ray source can
be seen in Figure \ref{fig:4649}(b) to have slight bulges along the axis of the radio source. These 
bulges appear to also be coincident with the axis of the 5 arcsec long central bar found by 
\citet{Randall04} in an analysis of \textit{Chandra} data. 
\citet{Randall04, Randall06} also suggest the existence of ``fingers'' in the 
X-ray emission which extend radially from the center of the galaxy. These features are on a much larger scale
than the radio emission, however, and thus are not present in the region shown in our images. We note 
that there are several small radio structures at the lowest contour level on the east and west sides of the 
X-ray source, some of which appear to coincide with depressions in the X-ray images. These are likely to be 
artifacts, though future observations may aid in determining whether or not they are real features.

\subsection{Jet Power}\label{sub:pjet}

In order to estimate the jet power required to evacuate the observed cavities in the X-ray emitting medium, we 
employ the method detailed in \citet{Allen06}. Using the radio contours as a guide, we model the cavities as 
ellipsoids with semi-axis dimensions $r_{\rm l}$ along the direction of the jet, $r_{\rm w}$ across it, and 
$r_{\rm d}$ along the line of sight. As the actual depth of the cavities is unknown, we assume symmetry about 
the axis defined by the direction of the jet, such that $r_{\rm d}=r_{\rm w}$. Thus the volume of the cavity is 
given by $V=4 \pi r_{\rm l} r_{\rm w}^2/3$, though systematic uncertainties in the cavity dimensions cause this 
value to be uncertain by about a factor of 2. This uncertainty in the volume is taken into account in  
subsequent calculations. The thermal pressure $P$ at the center of the cavity is determined from temperature 
and density measurements of the X-ray emitting gas. From these values, we can calculate the $P\,\textrm{d}V$ 
work done on the X-ray gas by the AGN jets in inflating the cavities. The energy $E$ required for this process 
is then found by adding the $P\,\textrm{d}V$ work done on the X-ray gas to the internal energy of the cavity:
\begin{equation}
E=P\,\textrm{d}V + \frac{1}{\gamma_2 - 1}PV = \frac{\gamma_2}{\gamma_2 - 1} PV. 
\end{equation}
Here $\gamma_2$ is the mean adiabatic index of the relativistic plasma contained in the cavity and has a value 
of 4/3. Thus the energy of the cavity is given by $E=4PV$.

The jet power can be estimated using the kinetic energy $E$ of the jet and the age of the cavity. We 
approximate the age of the cavity as the sound-speed expansion timescale $t_{\rm cs}$, which is the time 
required for the cavity to expand to its observed size at the speed of sound in its current environment (see 
e.g. \citealp{Dunn05, Birzan04}):
\begin{equation}
t_{\rm cs} = r_{\rm l}/c_s,
\end{equation}
where $c_s$ is the local sound speed. The jet power can then be estimated as
\begin{equation}
P_{\rm jet,\ cs} = E/t_{\rm cs}.
\end{equation}

Because the cavities consist of low density plasma and are thus expected to rise buoyantly through the 
surrounding medium, the age of the cavity can also be approximated as the buoyancy rise time $t_{\rm buoy}$, 
which is the time required for the cavity to rise buoyantly from the center of the galaxy to its current 
location. The buoyancy velocity is $v_{\rm buoy}=\sqrt{2gV/SC_{\rm D}}$, where $S=\pi r_{\rm w}^2$ is the 
cross-sectional area of the cavity, $C_{\rm D} \sim 0.75$ is the drag coefficient, and the gravitational 
acceleration is given by $g=GM(<R)/R^2$. The age, then, is approximately 
\begin{equation}
t_{\rm buoy}=R/v_{\rm buoy},
\end{equation}
where $R$ is the distance from the center of the black hole to the center of the cavity. The mass profiles 
used in the calculation of $v_{\rm buoy}$ come from the deprojection of the X-ray data assuming hydrostatic
equilibrium. The jet power can be calculated as 
\begin{equation}
P_{\rm jet,\ buoy} = E/t_{\rm buoy}. 
\end{equation}
The $P_{\rm jet}$ calculations are summarized in Table \ref{tab:pjet}.

\begin{table*}
\caption{Summary of the calculation of jet power $P_{\rm jet}$ (in units of $10^{41}$ erg s$^{-1}$). ``Lobe''
refers to the lobe for which the calculation is done (N and S for northern and southern lobe, respectively).
$R$ is the distance (in kpc) from the center of the black hole to
the center of the cavity, $P$ is the thermal pressure (in eV cm$^{-3}$) of the X-ray gas at radius $R$, $V$ is
the volume (in cm$^3$) of the cavity, $E$ is the kinetic energy (in units of $10^{54}$ erg) of the jets, 
$t_{\rm cs}$ is the sound speed expansion timescale (in $10^6$ yr), and $t_{\rm buoy}$ is the buoyancy rise time
(in $10^6$ yr).\label{tab:pjet}}
\begin{tabular}{lccccccccc}
\hline
\hline
Source & Lobe & $R$ & $P$ & $V$ & $E$ & $t_{\rm cs}$ & $t_{\rm buoy}$ & $P_{\rm jet,\ cs}$ & $P_{\rm jet,\ buoy}$ \\
\ & \ & (kpc) & (eV cm$^{-3}$) & ($10^{63}$ cm$^3$) & ($10^{54}$ erg) & ($10^6$ yr) & ($10^6$ yr) & ($10^{41}$ erg s$^{-1}$) & ($10^{41}$ erg s$^{-1}$) \\
\hline
NGC\,1399 & N & $8.4\pm0.8$ & $16.0\pm2.64$ & $1300\pm535$ & $131\pm59.2$ & $5.08\pm0.55$ & $12.6\pm3.83$ &  $8.24\pm3.70$ & $3.24\pm1.49$ \\
\ & S & $8.4\pm0.8$ & $16.0\pm2.65$ & $2950\pm1210$ & $300\pm134$ & $6.98\pm0.76$ & $10.8\pm3.30$ & $13.7\pm6.09$ & $8.57\pm3.93$ \\
NGC\,4649 & N & $1.1\pm0.1$ & $133\pm28.3$ & $50.0\pm20.4$ & $41.8\pm19.7$ & $1.91\pm0.20$ & $1.32\pm0.36$ & $6.95\pm3.26$ & $9.95\pm4.54$ \\
\ & S & $1.3\pm0.1$ & $134\pm28.3$ & $43.7\pm18.0$ & $36.6\pm17.5$ & $1.94\pm0.20$ & $1.72\pm0.48$ & $6.03\pm2.84$ & $6.63\pm3.11$ \\
\hline
\end{tabular}
\end{table*}

\subsection{A Note on the Timescales}\label{sub:timescales}

We have considered two different timescales in the calculations of jet power: one derived from the local sound
speed
and the other derived from the buoyancy velocity. The sound speed timescale measures the expansion of the cavity
at its current location, and is an appropriate estimate in cases in which the cavities are active, meaning that
they are currently powered by the jets of the AGN. The buoyancy timescale, however, measures the rise time
of the cavity to its current location, and is an appropriate estimate in the case of cavities that have detached
from the core and are rising buoyantly through the surrounding medium. Such is the case for ghost cavities,
which are not associated with GHz radio emission.

In the case of NGC\,1399, $t_{\rm buoy}$ is on average a factor of 2 greater than $t_{\rm cs}$. The cavities
are fairly well detached from the core, as can be seen in Figure \ref{fig:1399}. However, they are associated with
1.4~GHz emission and thus are presently 
powered by the AGN jets. Because the cavities in NGC\,1399 are detached but still active, it is
difficult to determine which timescale is the most appropriate description for the ages of the cavities in this
system. In NGC\,4649, the resolution of the radio image does not allow us to determine if the cavities are detached or
not, though both timescales are generally in agreement with each other within the errors.

\section{Particle Energies and Volume Filling Fractions}\label{sec:kf}

Using the synchrotron emission from the radio lobes, the energy contained within the relativistic electrons 
can be calculated.  Under the assumption that the radio lobes are in pressure balance with their surroundings, 
their particle content can be investigated. \citet{Fabian02} studied the lobes of 3C\,84 in the Perseus 
cluster, and subsequently \citet{Dunn04, Dunn05} investigated a large number of radio sources in galaxy 
clusters.  For a more detailed description of the method, see \citet{Dunn04}. Here we outline the important 
details and any changes to that method.

To calculate $E_{\rm e}$, the energy in the relativistic electrons, we assume a continuous synchrotron spectrum 
with a single spectral index $\alpha$, defined from $S_{\nu}\propto\nu^{\alpha}$, between $\nu_1=10$~MHz and 
$\nu_2=10$~GHz:
\begin{equation}\label{eq:Ee}
\begin{array}{l@{\hspace{0.05in}}l}
E_{\rm e} & = 4\pi\times10^{12} \left(\frac{cz}{H_0} \right)^2 \left(1+ \frac{z}{2} \right)^2 \frac{S_{\nu}}{\nu^{\alpha}} \frac{\nu_2^{0.5+\alpha}-\nu_1^{0.5+\alpha}}{\alpha+0.5} B^{-3/2} \\
\ & \approx aB^{-3/2} \\
\end{array}
\end{equation}
(e.g. \citealp{Fabian02}). We adopt spectral index values of $\alpha=-0.6$ and $\alpha=-0.7$ for the northern and 
southern lobes of NGC\,4649, respectively \citep{Stanger86}, and $\alpha=-0.92$ for NGC\,1399 \citep{Ekers89}. 
Also taking into account the energy within the magnetic field, the total energy in the lobes is given by
\begin{equation}
E_{\rm tot}=kE_{\rm e} + Vf\frac{B^2}{8\pi},
\end{equation}
where $k$ accounts for any other particles present in the lobe which are not accounted for by the simplistic 
model spectrum, $V$ is the volume of the cavity, and $f$ is the volume filling fraction of the relativistic 
plasma.

The magnetic field present inside the cavity is estimated by comparing the synchrotron cooling time of the 
plasma to the age of the lobe. The latter can be estimated from the sound-speed expansion timescale or the 
buoyancy rise time.  Under the assumption that the lobes are in pressure balance with their surroundings, we 
can obtain upper limits on the ratio $k/f$:  
\begin{equation}
\frac{k}{f}=\left( P-\frac{B^2}{8\pi}\right)\frac{3V}{a}B^{3/2}.
\end{equation}
If $k/f=1$, the lobe is uniformly
filled with a purely electron-positron plasma with energies corresponding to emission 
only in the range of $10\mhz$ to $10\ghz$.  If $f\sim 1$ and $k/f>1$, then $k>1$, which implies that there are 
``extra'' particles present.  These could be thermal protons mixed into the relativistic plasma as the jet 
travels out from the AGN, or they could be electrons which radiate outside of the assumed spectrum.

We calculate the $k/f$ values using timescales derived from the sound speed, as well as those derived from the
buoyancy velocity. These values are presented in Table \ref{tab:kf}. The equipartition values $k/f_{\rm eq}$ 
are calculated under the 
assumption that the plasma within the cavity is at equipartition with the pressure from the surrounding medium.
We use a Monte-Carlo algorithm to calculate the uncertainties in the upper limits obtained on $k/f$. The 
resulting distribution of limits on $k/f$ is non-gaussian.  We find that a log-normal distribution is a fair 
description of the data, and it is from this that the values for the uncertainties quoted in Table \ref{tab:kf}
are taken. The values for the uncertainties are large, approximately a factor of 3 of the $k/f$ values.

\begin{table*}
\caption{$k/f$ values. $S$ is the radio flux (in Jy) of the specified lobe, $k/f_{\rm eq}$ is the equipartition 
value, $k/f_{\rm sound}$ is calculated using the sound-speed expansion timescale, and $k/f_{\rm buoy}$ is 
calculated using the buoyancy rise time.
\label{tab:kf}}
\begin{tabular}{lccccc}
\hline
\hline
Galaxy & Lobe & $S$ (Jy) & $k/f_{\rm eq}$ & $k/f_{\rm sound}$ & $k/f_{\rm buoy}$ \\
\hline
NGC\,1399 & N & $0.11\pm0.01$ & $67.0_{-23.6}^{+190}$ & $93.0_{-30.3}^{+286}$ & $93.6_{-32.2}^{+273}$ \\
\ & S & $0.13\pm0.01$ & $129_{-45.8}^{+366}$  & $190_{-68.4}^{+531}$  & $177_{-59.0}^{+530}$ \\
NGC\,4649 & N & $0.0038\pm0.0002$ & $12000_{-4340}^{+33300}$ & $15800_{-6220}^{+40300}$ & $13600_{-3880}^{+47900}$ \\
\ & S & $0.0034\pm0.0002$ & $9450_{-3320}^{+26900}$  & $12400_{-4690}^{+32500}$ & $12000_{-3630}^{+39800}$ \\
\hline
\end{tabular}
\end{table*}

There is a striking difference between the $k/f$ limits, namely that those of NGC\,1399 are much lower than 
those of NGC\,4649.  This difference is linked to the radio flux, which is roughly a factor of 20 lower in 
NGC\,4649. The lower the value of $k/f$, the 
closer the assumed electron energy spectrum is to the actual one in the lobe.  However, as the synchrotron 
plasma ages, the high energy electrons lose their energy more rapidly, and the distribution of energies shifts 
to lower energies and also spreads out.  Therefore, observing old radio lobes at high frequencies results in 
low radio fluxes, which, if the spectrum is assumed to be flat, underestimates the number of particles present.
As a result, there would be a mismatch between the internal and external pressures.  In these calculations we 
force the lobes to be in pressure balance by increasing $k$ and/or decreasing $f$.

We currently have a lower cut-off to the synchrotron spectrum of
$\nu=10$~MHz.  However, in old lobes there would be very low energy
electrons, and so this lower cut-off might be too high.  Under the
assumption that $k/f=1$ and that we have correctly modeled the
synchrotron spectrum, we calculate the appropriate $\nu_1$.  We set an
absolute lower limit of the cyclotron frequency appropriate to the
magnetic field determined within the radio lobes.  For the northern
and southern lobes in NGC\,1399, $k/f=1$ can be obtained using a minimum
frequency of 265~Hz and 49~Hz respectively (the cyclotron frequencies
are 54~Hz and 43~Hz).  However, in NGC\,4649, even with the lower cut
off at the cyclotron frequency ($\sim 100$~Hz), $k/f$ is 2700 and
890 for the northern and southern lobes respectively.

Another explanation is that the assumed spectrum is close to what
exists in the cavity but that there are non-relativistic particles
present in the lobes, especially in NGC\,4649.  Thermal protons could
be entrained as the jet travels through the inner parts of the galaxy
and halo.  These would exert a pressure on the surrounding X-ray gas,
but would be undetectable from the radio synchrotron emission.

In NGC\,1399, the limits on $k/f$ implies that there are few entrained
particles and that the assumed spectrum does describe the distribution
of electrons in the lobe reasonably well.  However, in NGC\,4649
either there are many entrained particles or the model spectrum does
not describe the electron population well.

There is close agreement between the
limits obtained from the sound speed and buoyancy timescales.  We find
that the only lobes which can be in equipartition between the leptons
and the magnetic field are those from NGC\,4649 if the buoyancy
timescale is used to calculate the magnetic field from the synchrotron 
cooling time.

\section{Conclusions}\label{sec:concl}

In analyzing \textit{Chandra} and VLA data for NGC\,4649, we have found cavities in the surrounding X-ray 
emitting gas that have not been previously reported. We used these cavities and those in NGC\,1399 to estimate 
the power of the jets emanating from the central black hole of these systems. The $P_{\rm jet}$ values for
NGC\,1399 using the sound speed expansion timescale 
are $\sim8\times10^{41}$ erg s$^{-1}$ for the northern lobe and $\sim14\times10^{41}$ erg s$^{-1}$
for the southern lobe. The values using the buoyancy rise time are on average a 
factor of 2 lower for both lobes, but are roughly in agreement with the aforementioned values within the 
errors. As the cavities in NGC\,1399 are detached from the core but presently powered by the AGN jets, it is
difficult to determine which timescale is the relevant one in this case. In NGC\,4649, the $P_{\rm jet}$ 
values using the sound speed expansion timescale are $\sim7\times10^{41}$ erg s$^{-1}$ for the northern lobe 
and $\sim6\times10^{41}$ erg s$^{-1}$ for the southern lobe. Calculations using the buoyancy rise time yield
similar values, within the errors. 

Comparing these jet powers
to those calculated in \citet{Allen06}, we find that our sources have jet powers that are lower
than most of those in their sample. Examples are M\,87 and NGC\,4696, both of which are located within 
dense cluster environments. M\,87 
is near the center of the Virgo cluster and has jet powers which are roughly an order of magnitude larger 
than those of our sources. The jet powers for the lobes of NGC\,4696, which is in the center of Centaurus, 
are roughly 3 to 4 times those of our sources. We find that our sources have similar jet powers to those of 
NGC\,4552 in Virgo, which has values of $\sim7\times10^{41}$ erg s$^{-1}$ and $\sim8\times10^{41}$ 
erg s$^{-1}$, as well as those of NGC\,5846, which dominates a group of intermediate mass and has 
jet powers of $\sim4\times10^{41}$ erg s$^{-1}$ for each lobe.

We also investigated the particle content of the cavities by calculating $k/f$. We find that the $k/f$ values 
for NGC\,4649 are much larger 
than those of NGC\,1399 by about 2 orders of magnitude. \citet{Dunn05} found that there is a large range of 
values for $k/f$ in cavities found in cluster galaxies. In general, $k/f$ is in the range of $\sim 1$ to 
$\sim 1000$
for cavities that are active. The values for NGC\,1399 fall into this range; however, in the case of NGC\,4649, 
the values for $k/f$ using either timescale are more than a factor of 10 larger than the upper limit of what 
has been previously observed. 

The minimum value of $k/f=1$ corresponds to a lobe which is filled with a uniform electron-positron plasma
with energies corresponding to emission in the range of 10~MHz to 10~GHz. 
Large values of $k/f$, such as those found in NGC\,4649, can be achieved by large values for $k$ or small
values for $f$. Large values for $k$ imply that, in order for there to be pressure balance, there must
be particles present within the lobe which are not accounted for by the assumed spectrum. These particles
can be electrons whose emission is outside the range of 10~MHz to 10~GHz or thermal protons that either exist
in the jets during their formation or are entrained as the cavities travel through the surrounding medium.
The filling factor $f$ describes the fraction of the volume of the cavity that is occupied
by radio emitting plasma. The remaining volume fraction can either be occupied 
by thermal gas or plasma whose relativistic electrons have aged. 
\citet{Dunn05}, in studying ghost 
cavities as well as active cavities in Centaurus and Perseus, found that the older cavities have larger $k/f$
values than the active ones and have attributed this to the aging of the relativistic electrons. 

The $k/f$ values for NGC\,4649 are similar to those of ghost cavities in which the relativistic electrons have 
aged, though the values can also suggest that the the lobes contain entrained particles.
If the electrons in the lobes of NGC\,4649 have aged, the model spectrum is simply not an accurate description of 
the electron distribution. This scenario seems likely, as the radio flux is so small that NGC\,4649 is 
expected to soon become a ghost cavity system should clear cavities remain.
For NGC\,1399, the model spectrum appears to be an appropriate description of the lobe contents, and there are 
probably not many entrained particles. As was proposed by \citet{Dunn04}, higher power radio sources are less likely
to pick up extra particles from the surrounding medium. Our results agree with this description in that NGC\,1399 
has lower $k/f$ values than
NGC\,4649 and also higher radio power. However, there are many physical parameters to consider in
investigating these values, and no clear correlation has been found between $k/f$ and any one parameter.

In calculating $k/f$ we have assumed a simple power-law spectrum. Using high frequency observations to obtain radio 
flux measurements of a lobe whose relativistic electrons have aged would tend to cause this power-law approximation 
to underestimate the energy of the relativistic electrons in the lobe 
and therefore overestimate $k/f$. Our results, then, could be said to suggest that 
the lobes in NGC\,4649 are likely to be older than those in NGC\,1399. However, these results could also suggest that 
the power-law approximation may be poor in both cases. In order to determine whether the power-law assumption is 
reasonable, it would be desirable to obtain sensitive lower frequency 
VLA observations of these galaxies, such as at 330~MHz, that would allow for accurate measurements of the
radio flux in the regions corresponding to the cavities.

\section*{Acknowledgements}
We would like to thank the anonymous reviewer for constructive suggestions.
We acknowledge support from the National Aeronautics and Space Administration through Chandra Award Number 
GO4-5134A issued by the Chandra X-ray Observatory Center, which is operated by the Smithsonian Astrophysical 
Observatory for and on behalf of the National Aeronautics and Space Administration under contract NAS8-03060. 
This research has made use of the NASA/IPAC Extragalactic Database (NED), which is operated by the Jet 
Propulsion Laboratory, Caltech, under contract with NASA. The National Radio Astronomy Observatory is a 
facility of the National Science Foundation operated under a cooperative agreement by Associated 
Universities, Inc.


\begin{thebibliography}{}

\bibitem[Allen et al.(2006)]{Allen06} Allen S.~W., Dunn R.~J.~H., Fabian A.~C., 
Taylor G.~B., Reynolds C.~S., 2006, MNRAS, 372, 21

\bibitem[Arnaud (1996)]{Arnaud96} Arnaud K.~A., 1996, in Jacoby G.~H., Barnes J.,
eds, ASP Conf. Ser.101: Astronomical Data Analysis Software and Systems V XSPEC:
The First Ten Years. p.17

\bibitem[Bernardi et al.(2006)]{Bernardi06} Bernardi M., Hyde J.~B., Sheth R.~K.,
Miller C.~J., Nichol R.~C., 2006, astro-ph/0607117

\bibitem[Best et al.(2005)]{Best05} Best P.~N., Kauffmann G., Heckman T.~M., 
Brinchmann J., Charlot S., Ivezi\'{c} \u{Z}., White S.~D.~M., 2005, MNRAS, 
362, 25-40

\bibitem[Best et al.(2006)]{Best06b} Best P.~N., von der Linden A., Kauffmann G., 
Heckman T.~M., Kaiser C.~R., 2006, astro-ph/0611197

\bibitem[Beuing et al.(1999)]{Beuing99} Beuing J., Dobereiner S., B\"{o}hringer H.,
Bender R., 1999, MNRAS, 302, 209

\bibitem[B\^{i}rzan et al.(2004)]{Birzan04} B\^{i}rzan L., Rafferty D.~A., 
McNamara B.~R., Wise M.~W., Nulsen P.~E.~J., 2004, ApJ, 607, 800 

\bibitem[B\"{o}hringer et al.(1993)]{Bohringer93} B\"{o}hringer H., Voges W., 
Fabian A.~C., Edge A.~C., Neumann D.~M., 1993, MNRAS, 264, L25

\bibitem[Bondi(1952)]{Bondi52} Bondi H., 1952, MNRAS, 112, 195

\bibitem[Churazov et al.(2001)]{Churazov01} Churazov E., Br\"{u}ggen M., Kaiser 
C.~R., B\"{o}hringer H., Forman W., 2001, ApJ, 554, 261

\bibitem[Condon et al.(1998)]{Condon98} Condon J.~J., Cotton W.~D., Greisen E.~W., 
Yin Q.~F., Perley R.~A., Taylor G.~B., Broderick J.~J., 1998, AJ, 115, 1693 

\bibitem[Di Matteo et al.(2003)]{DiMatteo03} Di Matteo T., Allen S.~W., Fabian 
A.~C., Wilson A.~S., Young A.~J., 2003, ApJ, 582, 133

\bibitem[Diehl \& Statler(2006)]{Diehl06a} Diehl S., Statler T.~S., 2006, 
astro-ph/0606215

\bibitem[Donahue et al.(2006)]{Donahue06} Donahue M., Horner D.~J., Cavagnolo K.~W.,
Voit G.~M., 2006, ApJ, 643, 730

\bibitem[Dunn \& Fabian(2004)]{Dunn04} Dunn R.~J.~H., Fabian A.~C., 2004, MNRAS, 
355, 862-873

\bibitem[Dunn, Fabian \& Taylor(2005)]{Dunn05} Dunn R.~J.~H., Fabian A.~C., Taylor G.~B., 2005,
MNRAS, 364, 1343

\bibitem[Dunn \& Fabian(2006)]{Dunn06f} Dunn R.~J.~H., \& Fabian A.~C., 2006, MNRAS,
373, 959

\bibitem[Ebeling et al.(2006)]{Ebeling06} Ebeling H., White D.~A.,  Rangarajan F.~V.~N.,
 2006, MNRAS, 368, 65 

\bibitem[Ekers et al.(1989)]{Ekers89} Ekers R.~D., Wall J.~V., Shaver P.~A., Goss W.~M., 
Fosbury R.~A.~E., Danziger I.~J., Moorwood A.~F.~M., Malin D.~F., Monk A.~S., Ekers J.~A.,
1989, MNRAS, 236, 737

\bibitem[Fabian et al.(2002)]{Fabian02} Fabian A.~C., Celotti A., Blundell K.~M., 
Kassim N.~E., Perley R.~A., 2002, MNRAS, 331, 369

\bibitem[Fabian et al.(2003)]{Fabian03b} Fabian A.~C., Sanders J.~S., Allen S.~W., 
Crawford C.~S., Iwasawa K., Johnstone R.~M., Schmidt R.~W., Taylor G.~B., 
2003, MNRAS, 344, L43

\bibitem[Fabian et al.(2000)]{Fabian00} Fabian A.~C., Sanders J.~S., Ettori S.,
Taylor G.~B., Allen S.~W., Crawford C.~S., Iwasawa K., Johnstone R.~M., Ogle
P.~M., 2000, MNRAS, 318, L65

\bibitem[Fabian et al.(2006)]{Fabian06} Fabian A.~C., Sanders J.~S., Taylor G.~B., 
Allen S.~W., Crawford C.~S., Johnstone R.~M., Iwasawa K., 2006, MNRAS, 366, 417

\bibitem[Fanaroff \& Riley(1974)]{Fanaroff74} Fanaroff B.~L. \& Riley J.~M., 1974, MNRAS,
167, 31P

\bibitem[Finoguenov \& Jones(2001)]{Finoguenov01} Finoguenov A., \& Jones C.,  2001,
\apjl, 547, L107 

\bibitem[Forman et al.(2005)]{Forman05} Forman W., Nulsen P., Heinz S., Owen F., 
Eilek J., Vikhlinin A., Markevitch M., Kraft R., Churazov E., Jones C., 2005,
ApJ, 635, 894

\bibitem[Mahdavi \& Geller(2001)]{Mahdavi01} Mahdavi A., Geller M.~J., 2001, ApJ, 554, L129

\bibitem[McNamara et al.(2000)]{McNamara00} McNamara B.~R., Wise M., Nulsen P.~E.~J., 
David L.~P., Sarazin C.~L., Bautz M., Markevitch M., Vikhlinin A., Forman W.~R., 
Jones C., Harris D.~E., 2000, ApJ, 534, L135

\bibitem[Mewe et al.(1995)]{Mewe95} Mewe R., Kaastra J.~S., Leidahl D.~A., 1995, Legacy,
6, 16

\bibitem[O'Sullivan et al.(2001)]{OSullivan01} O'Sullivan E., Forbes D.~A., Ponman T.~J.,
2001, MNRAS, 328, 461

\bibitem[Paolillo et al.(2002)]{Paolillo02} Paolillo M., Fabbiano G., Peres G., Kim
D.-W., 2002, ApJ, 565, 883

\bibitem[Randall et al.(2004)]{Randall04} Randall S.~W., Sarazin C.~L., Irwin J.~A.,
2004, ApJ, 600, 729

\bibitem[Randall et al.(2006)]{Randall06} Randall S.~W., Sarazin C.~L., Irwin J.~A.,
2006, ApJ, 636, 200

\bibitem[Rafferty et al.(2006)]{Rafferty06} Rafferty D.~A., McNamara B.~R., Nulsen 
P.~E.~J., Wise M.~W., 2006, ApJ, 652, 216

\bibitem[Stanger \& Warwick(1986)]{Stanger86} Stanger V.~J., \& Warwick, R.~S.,
1986, MNRAS, 220, 363

\bibitem[Taylor et al.(2002)]{Taylor02} Taylor G.~B., Fabian A.~C., Allen S.~W.,
2002, MNRAS, 334, 769

\bibitem[Taylor et al.(2006)]{Taylor06a} Taylor G.~B., Sanders J.~S., Fabian A.~C., 
Allen S.~W., 2006, MNRAS, 365, 705

\bibitem[Tremaine et al.(2002)]{Tremaine02} Tremaine S., Gebhardt K., Bender R., 
Bower G., Dressler A., Faber S.~M., Filippenko A.~V., Green R., Grillmair C., 
Ho L.~C., Kormendy J., Lauer T.~R., Magorrian J., Pinkney J., Richstone D., 
2002, ApJ, 574, 740

\bibitem[Wegner et al.(2003)]{Wegner03} Wegner G., Bernardi M., Willmer C.~N.~A.,
da Costa L.~N., Alonso M.~V., Pellegrini P.~S., Maia M.~A.~G., Chaves O.~L.,
Rit\'{e} C., 2003, AJ, 126, 2268

\end{thebibliography}
\end{document}